\documentclass[twocolumn]{aastex63}
\usepackage{upgreek}

\shorttitle{}
\shortauthors{}

\begin{document}

\title{SCExAO/MEC and CHARIS Discovery of a Low Mass, 6 AU-Separation Companion to HIP 109427 using Stochastic Speckle Discrimination and High-Contrast Spectroscopy\footnote{Based in part on data collected at Subaru Telescope, which is operated by the National Astronomical Observatory of Japan.}}

\correspondingauthor{Sarah Steiger}
\email{steiger@ucsb.edu}

\author[0000-0002-4787-3285]{Sarah Steiger}
\affiliation{Department of Physics, University of California, Santa Barbara, Santa Barbara, California, USA}
\author[0000-0002-7405-3119]{Thayne Currie}
\affiliation{Subaru Telescope, National Astronomical Observatory of Japan, 
650 North A`oh$\bar{o}$k$\bar{u}$ Place, Hilo, HI  96720, USA}
\affiliation{NASA-Ames Research Center, Moffett Blvd., Moffett Field, CA, USA}
\affiliation{Eureka Scientific, 2452 Delmer Street Suite 100, Oakland, CA, USA}
\author{Timothy D. Brandt}
\affiliation{Department of Physics, University of California, Santa Barbara, Santa Barbara, California, USA}
\author{Olivier Guyon}
\affiliation{Subaru Telescope, National Astronomical Observatory of Japan, 
650 North A`oh$\bar{o}$k$\bar{u}$ Place, Hilo, HI  96720, USA}
\affil{Steward Observatory, The University of Arizona, Tucson, AZ 85721, USA}
\affil{College of Optical Sciences, University of Arizona, Tucson, AZ 85721, USA}
\affil{Astrobiology Center, 2-21-1, Osawa, Mitaka, Tokyo, 181-8588, Japan}
\author{Masayuki Kuzuhara}
\affiliation{Astrobiology Center, 2-21-1, Osawa, Mitaka, Tokyo, 181-8588, Japan}
\affiliation{National Astronomical Observatory of Japan, 2-21-2, Osawa, Mitaka, Tokyo 181-8588, Japan}
\author{Jeffrey Chilcote}
\affiliation{Department of Physics, University of Notre Dame, South Bend, IN, USA}
\author{Tyler D. Groff}
\affiliation{NASA-Goddard Space Flight Center, Greenbelt, MD, USA}
\author{Julien Lozi}
\affiliation{Subaru Telescope, National Astronomical Observatory of Japan, 
650 North A`oh$\bar{o}$k$\bar{u}$ Place, Hilo, HI  96720, USA}
\author{Alexander B. Walter}
\affiliation{Jet Propulsion Laboratory, California Institute of Technology, Pasadena, California 91125, USA}
\author{Neelay Fruitwala}
\affiliation{Department of Physics, University of California, Santa Barbara, Santa Barbara, California, USA}
\author[0000-0002-4272-263X]{John I. Bailey, III}
\affiliation{Department of Physics, University of California, Santa Barbara, Santa Barbara, California, USA}
\author[0000-0003-3146-7263]{Nicholas Zobrist}
\affiliation{Department of Physics, University of California, Santa Barbara, Santa Barbara, California, USA}
\author[0000-0001-5721-8973]{Noah Swimmer}
\affiliation{Department of Physics, University of California, Santa Barbara, Santa Barbara, California, USA}
\author[0000-0003-4792-6479]{Isabel Lipartito}
\affiliation{Department of Physics, University of California, Santa Barbara, Santa Barbara, California, USA}
\author[0000-0002-0849-5867]{Jennifer Pearl Smith}
\affiliation{Department of Physics, University of California, Santa Barbara, Santa Barbara, California, USA}
\author{Clint Bockstiegel}
\affiliation{CERN - 1211 Geneva 23 - Switzerland}
\author{Seth R. Meeker}
\affiliation{Jet Propulsion Laboratory, California Institute of Technology, Pasadena, California 91125, USA}
\author{Gregoire Coiffard}
\affiliation{Department of Physics, University of California, Santa Barbara, Santa Barbara, California, USA}
\author{Rupert Dodkins}
\affiliation{Department of Physics, University of California, Santa Barbara, Santa Barbara, California, USA}
\author{Paul Szypryt}
\affiliation{National Institute of Standards and Technology, Boulder, Colorado 80305, USA}
\affiliation{Department of Physics, University of Colorado, Boulder, Colorado 80309, USA}
\author[0000-0001-5587-845X]{Kristina K. Davis}
\affiliation{Department of Physics, University of California, Santa Barbara, Santa Barbara, California, USA}
\author{Miguel Daal}
\affiliation{Department of Physics, University of California, Santa Barbara, Santa Barbara, California, USA}
\author{Bruce Bumble}
\affiliation{Jet Propulsion Laboratory, California Institute of Technology, Pasadena, California 91125, USA}
\author{Sebastien Vievard}
\affiliation{Subaru Telescope, National Astronomical Observatory of Japan, 
650 North A`oh$\bar{o}$k$\bar{u}$ Place, Hilo, HI  96720, USA}
\author{Ananya Sahoo}
\affiliation{Subaru Telescope, National Astronomical Observatory of Japan, 
650 North A`oh$\bar{o}$k$\bar{u}$ Place, Hilo, HI  96720, USA}
\author{Vincent Deo}
\affiliation{Subaru Telescope, National Astronomical Observatory of Japan, 
650 North A`oh$\bar{o}$k$\bar{u}$ Place, Hilo, HI  96720, USA}
\author[0000-0001-5213-6207]{Nemanja Jovanovic}
\affiliation{Department of Astronomy, California Institute of Technology, 1200 E. California Blvd.,Pasadena, CA, 91125, USA}
\author{Frantz Martinache}
\affiliation{Universit\'{e} C\^{o}te d'Azur, Observatoire de la C\^{o}te d'Azur, CNRS, Laboratoire Lagrange, France}
\author{Greg Doppmann}
\affil{W.M. Keck Observatory, HI, USA}
\author{Motohide Tamura}
\affil{Astrobiology Center, 2-21-1, Osawa, Mitaka, Tokyo, 181-8588, Japan}
\affiliation{National Astronomical Observatory of Japan, 2-21-2, Osawa, Mitaka, Tokyo 181-8588, Japan}
\affiliation{Department of Astronomy, Graduate School of Science, The University of Tokyo, 7-3-1, Hongo, Bunkyo-ku, Tokyo, 113-0033, Japan}
\author{N. Jeremy Kasdin}
\affiliation{University of San Francisco, San Francisco, CA 94118}
\author[0000-0003-0526-1114]{Benjamin A. Mazin}
\affiliation{Department of Physics, University of California, Santa Barbara, Santa Barbara, California, USA}

\begin{abstract}
We report the direct imaging discovery of a low-mass companion to the nearby accelerating A star, HIP 109427, with the Subaru Coronagraphic Extreme Adaptive Optics (SCExAO) instrument coupled with the MKID Exoplanet Camera (MEC) and CHARIS integral field spectrograph. CHARIS data reduced with reference star PSF subtraction yield 1.1--2.4~$\upmu$m spectra. MEC reveals the companion in $Y$ and $J$ band at a comparable signal-to-noise ratio using stochastic speckle discrimination, with no PSF subtraction techniques. Combined with complementary follow-up $L_{\rm p}$ photometry from Keck/NIRC2, the SCExAO data favors a spectral type, effective temperature, and luminosity of M4--M5.5, 3000-3200 $K$, and $\log_{10}(L/L_{\rm \odot}) = -2.28^{+0.04}_{-0.04}$, respectively. Relative astrometry of HIP 109427 B from SCExAO/CHARIS and Keck/NIRC2, and complementary Gaia-Hipparcos absolute astrometry of the primary favor a semimajor axis of $6.55^{+3.0}_{-0.48}$ au, an eccentricity of $0.54^{+0.28}_{-0.15}$, an inclination of $66.7^{+8.5}_{-14}$ degrees, and a dynamical mass of $0.280^{+0.18}_{-0.059}$ $M_{\odot}$.   This work shows the potential for extreme AO systems to utilize speckle statistics in addition to widely-used post-processing methods to directly image faint companions to nearby stars near the telescope diffraction limit.
\pagebreak
\end{abstract}



\section{Introduction} \label{sec:intro}

Nearly all of the $\sim$ 10--20 directly imaged planets discovered so far orbit their host stars at 10--150 au separations, typically $\rho$ $\sim$ 0\farcs{}4--2\arcsec{} on the sky \citep[e.g.][]{Marois2008a,Lagrange2009, Rameau2013, kuzuhara2013, Currie2014a,macintosh2015,Chauvin2017}. The first generation of extreme adaptive optics (AO) instruments, such as the Gemini Planet Imager \citep[GPI;][]{macintosh2014} and the Spectro-Polarimetric High-contrast Exoplanet REsearch at VLT \citep[SPHERE;][]{beuzit2019}, have achieved factors of 100 improvement in contrast at sub-arcsecond separations over conventional systems, but typically were only sensitive to jovian exoplanets at projected separations beyond $\sim$ 10 au \citep[e.g.][]{Nielsen2019,Vigan2020}.  To more frequently identify companions at Jupiter-to-Saturn separations, upgraded versions of GPI/SPHERE and second-generation systems like SCExAO and MagAO-X \citep{Jovanovic2015,Males2020} must yield deeper contrasts at $\rho$ $<$ 0\farcs{}4.

Point spread function (PSF) sized speckles with a range of correlation timescales ($\tau$) and sources currently limit achievable contrasts from the ground. Rapidly-evolving atmospheric speckles ($\tau$ $\sim$ 1-20 ms) result from aberrations left uncorrected by an AO system and average out over the course of long-exposure images, forming a smooth halo \citep[e.g.][]{Perrin2003,Soummer2007}. These ``fast" speckles can be corrected by improved AO control loops which will mitigate temporal bandwidth error and measurement (photon noise) error \citep[e.g.][]{Guyon2005}. Alternatively, quasi-static speckles result from imperfections in the instrument such as non-common path errors, telescope vibrations etc. \citep{Guyon2005, lozi2018}. These speckles interfere with atmospheric speckles and can be pinned to the diffraction rings \citep{Soummer2007}.  Quasi-static speckle noise follows a highly non-Gaussian (modified Rician distribution) and is temporally well correlated ($\tau$ $\sim$ 10-60 minutes), presenting a fundamental obstacle in exoplanet direct imaging \citep[e.g.][]{Marois2008b}.    

While focal-plane wavefront control methods can conceivably suppress these speckles \citep[e.g.][]{Give'on2007, martinache2016}, post-processing methods provide the most common way of removing them.   Unfortunately, common post-processing techniques utilizing advanced PSF subtraction methods \citep[e.g.][]{Lafreniere2007,Soummer2012} become less effective at small angles where direct detections are most challenging. Angular Differential Imaging \citep[ADI;][]{Marois2006} exploits parallactic angle (PA) rotation to distinguish speckles, which will rotate with the telescope field of view, from companions, which are at a fixed location on-sky. The magnitude of this rotation, however, scales proportionally with angular separation for a given unit time, resulting in less rotation at smaller inner working angles (IWAs).  Additionally, the rotation in $\lambda$/D units is smaller within a few diffraction beamwidths, resulting in severe self-subtraction of a planet signal \citep{mawet2012}. Similarly, Spectral Differential Imaging \citep[SDI;][]{Marois2000} utilizes the wavelength-independent nature of phase-induced speckle noise to rescale (magnify) slices of polychromatic images.   However, SDI requires broad spectral coverage close to the primary otherwise it also suffers from self-subtraction effects. Reference Star Differential Imaging \citep[RDI/RSDI;][]{Soummer2012} does not inherently suffer at small IWA, but requires careful magnitude and color matching between the target of interest and the reference star. Mismatches in the direction of the gravity vector with respect to the primary mirror and in the position of the telescope rotator between reference observations and target observations can also degrade RDI performance, placing even tighter constraints on the choice of reference star \citep{ruane2019}. A method to suppress quasi-static speckles that is free of the limitations of ADI, SDI, and RDI would significantly improve our ability to detect jovian planets at Jupiter-to-Saturn like separations.

Here we demonstrate the use of a post-processing technique called Stochastic Speckle Discrimination \citep[SSD;][]{fitzgerald2006speckle, gladysz2008, meeker2018darkness} for detecting new low mass companions using SCExAO and the Microwave Kinetic Inductance Detector (MKID) Exoplanet Camera \citep[MEC;][]{walter2020mkid}. SSD works by utilizing the timing resolution of MKID detectors to break up an observation into a series of short exposures in post-processing. These short exposure images are then used to generate intensity histograms for each pixel in an image. If the time binning is short enough, we can adequately sample the underlying probability density function (PDF) that describes the off-axis intensity in the image (light from a speckle) which can be written analytically as a modified Rician distribution. Fitting this distribution to the intensity histograms then allows us to diagnose whether a bright point in an image is a quasi-static speckle, or a true companion, see Section \ref{sec:SSD}.

We also report the discovery of a low mass stellar companion to HIP 109427 using, in part, SSD with SCExAO/MEC. We also utilize SCExAO/MEC photometry, SCExAO/CHARIS spectroscopy, and Keck/NIRC2 photometry. This companion has a best fit dynamical mass of $\sim$ 0.25 $M_{\odot}$ consistent with a spectral type of M4--M5.5 from spectral analysis.   



This discovery serves as an important proof-of-concept for the use of time-domain information in addition to standard PSF subtraction methods exploiting spectral and spatial information to remove quasi-static speckles in high-contrast images.

 \begin{deluxetable*}{llllllllll}
    \tablewidth{0pt}
    \tabletypesize{\scriptsize}
    \tablecaption{HIP 109427 Observing Log}
    \tablehead{\colhead{UT Date} & \colhead{Instrument} &  \colhead{coronagraph} & \colhead{Seeing (\arcsec{})} &{Passband} & \colhead{$\lambda$ ($\mu m$)$^{a}$} 
    & \colhead{$t_{\rm exp}$ (s)} & \colhead{$N_{\rm exp}$} & \colhead{$\Delta$PA ($^{o}$)} & \colhead{Post-Processing} \\
    {} & {} & {} & {} & {} & {} & {} & {} & {} & \colhead{Strategy}  }
    \startdata
    \textbf{New Data}\\
    20200731 & SCExAO/CHARIS & Lyot & 0.6 & $JHK$ & 1.16--2.37 &10.32 & 43 & 5.4 & RDI/KLIP\\
    20201007 & SCExAO/MEC & Lyot & 0.35 & $YJ$ & 0.95--1.14 &25.0 & 36 & 2.3 & SSD\\
     -- & SCExAO/CHARIS & Lyot & 0.35 & $H$ & 1.48--1.79 &16.23-20.65$^{b}$ & 78 & 5.4 & none\\
     20201225 & Keck/NIRC2+PyWFS & none & 0.7 & $Lp$ & 3.78 &22.5 & 49 & 3.5 & RDI/KLIP\\
     \textbf{Archival Data}\\
     20151028 & Keck/NIRC2 & vortex & 0.7 & $Lp$ & 3.78 & 25 & 25 & 11.6 & RDI/ALOCI
    \enddata
    \tablecomments{a) For CHARIS and MEC data, this column refers to the wavelength range.  For broadband imaging data, it refers to the central wavelength. b) Total integration time is 1524 $s$.
    }
    \label{obslog_hip109427}
    \vspace{-0.5cm}
\end{deluxetable*}

\section{System Properties and Observations} 
 HIP 109427 (tet Peg) is a nearby ($d$ = 28.3 pc) $\lambda$ Boo star with a spectral type of A1V \citep{Gray2006, vanLeeuwen2007}.    \citet{David2015} and \citet{Stone2018} derive system ages of $t$ $\sim$ 400--700 Myr; Banyan-$\Sigma$ does not reveal evidence that the star's kinematics are consistent with younger moving groups \citep{Gagne2018}.   While HIP 109427 lacks a published detected radial-velocity trend indicative of a companion \citep{Lagrange2009,Howard2016}, \citet{Makarov2005} suggest evidence for a potential companion at a 5.7$\sigma$ level from Hipparcos astrometry. Previous direct imaging observations taken as a part of the thermal infrared LEECH survey conducted with the Large Binocular Telescope failed to image any companions \citep{Stone2018}.   Searches through public archives show that the star has not been targeted as a part of the Gemini Planet Imager campaign planet search, but it has been observed with VLT/NaCo and SPHERE without a reported companion. 
 
 Astrometry derived from the \textit{Hipparcos-Gaia Catalogue of Accelerations} \citep[HGCA;][]{brandt2018hipparcos} reveals a substantial deviation from simple linear kinematic motion ($\chi^{2}$ = 108.83) consistent with a $\sim$ 11-$\sigma$-significant acceleration. We therefore targeted this star as a part of our survey to discover low-mass companions to accelerating stars \citep[e.g.][]{Currie2020}.

 In three epochs between July and December 2020, we observed HIP 109427 with the Subaru Telescope using SCExAO coupled to CHARIS and MEC and with the Keck II telescope using the NIRC2 camera. \citep{Jovanovic2015,Groff2016, Currie2020b,walter2020mkid} (Table \ref{obslog_hip109427}).
Conditions were photometric each night with average to excellent optical seeing ($\theta_{\rm V}$ = 0\farcs{}35--0\farcs{}7).    

The SCExAO Pyramid wavefront sensor (PyWFS) ran at 2 kHz, correcting for 1080 spatial modes and delivering a diffraction-limited PSF core.   
MEC data (7 October 2020) covers wavelengths over the $Y$  and $J$ passbands (0.95 - 1.4 $\upmu$m)  at a spectral resolution of $\mathcal{R}$ $\sim$ 3.3. 
We obtained CHARIS data in broadband (1.1--2.4 $\upmu$m; 31 July 2020) at a resolution of $\mathcal{R}$ $\sim$ 18 or in $H$ band at a higher resolution ($\mathcal{R}$ $\sim$ 70).

The Keck near-IR PyWFS \citep{Bond2020} corrected the wavefront at 1 kHz, correcting for 349 spatial modes and NIRC2 data (25 December 2020) was taken in the $L_{\rm p}$ broadband filter ($\lambda_{o}$ = 3.78~$\upmu$m).


All observations were conducted in ``vertical angle''/pupil tracking mode, enabling angular differential imaging \citep[ADI;][]{Marois2006}.  The CHARIS data also enables spectral differential imaging \citep[SDI;][]{Marois2000}.
 CHARIS and MEC data utilized the Lyot coronagraph  (0\farcs{}23 diameter) to suppress the stellar halo, as well as satellite spots for precise astrometric and spectrophotometric calibration \citep[e.g.][]{Jovanovic2015-astrogrids,Currie2018, sahoo2020}. NIRC2 exposures left the HIP 109427 primary unocculted and unsaturated.   Parallactic angle rotation for all data sets was small to negligible; however, we obtained  reference star observations for the CHARIS broadband and NIRC2 data (HIP 105819 and HIP 112029, respectively).   

 To complement these new data, we analyzed Keck/NIRC2 $L_{\rm p}$ data for HIP 109427 taken on 28 October 2015 from the Keck Observatory Archive (Program ID C197NI).   These data were obtained with Keck II's facility (Shack-Hartmann) adaptive optics system and the vector vortex coronagraph \citep{Serabyn2017}.   We used HD 212061, observed immediately after HIP 109427, for reference star subtraction.
 \section{Data}

 \subsection{Image Processing: MEC}
 \subsubsection{Basic Processing}
    
MEC data was reduced using the MKID Data Reduction and Analysis Pipeline \citep{walter2020mkid}\footnote{GitHub: \url{https://github.com/MazinLab/MKIDPipeline}}. This pipeline notably includes a wavelength calibration, a flat-field correction, and a spectrophotometric calibration amongst other steps.  The MKID Pipeline can output calibrated images in a fits file format to be able to interface with traditional post processing techniques and astronomical image viewing software, but can also output microsecond precision, time-tagged photon lists due to the unique nature of MKID detectors.

Each pixel in an MKID array is a superconducting LC resonant circuit with a photosensitive inductor and tunable interdigitated capacitor. When a photon strikes the inductor of the MKID pixel, cooled below its transition temperature, quasiparticles are generated by the breaking of Cooper pairs. This increases the inductance of the material and lowers the resonant frequency of the circuit. This is analogously measured as a change in phase by room temperature readout electronics. Since the number of quasiparticles generated is proportional to the energy of the incident photon, MKIDs have an inherent energy resolution without the use of filters or gratings \citep{day2003, mazin2012superconducting, szypryt2017large}. Additionally, each resonator is sampled at a rate of 1 MHz, yielding microsecond timing resolution \citep{fruitwala2020}.

This precise timing information makes MKID instruments like MEC well suited to perform time-domain based post processing techniques like Stochastic Speckle Discrimination (SSD) as described below.

As with the CHARIS data, satellite spots were used for the spectrophotometric calibration reference. We adopted the scaling between modulation amplitude and contrast from \citet{Currie2018b} to generate the expected satellite spot flux values per passband.   A stellar spectrum from the PHOENIX stellar library appropriate for an A1V star was used and the data normalized to match HIP 109427's reported J band flux \citep{Ducati2002}. Given MEC's low energy resolution, we focused on broadband MEC photometry (not spectra). Additionally, due to the wavelength scaling of the spots, the satellite spots are extended out into elongated streaks instead of appearing as copies of an unocculted stellar PSF. This is similar to the case for GPI's polarimetry mode.   

To derive photometry for the satellite spots, we therefore follow similar methods to those outlined for GPI's polarimetry mode from \citet{Millar-Blanchaer2016}.   Briefly, we subtract off a plane fitted background from a region surrounding each of the four satellite spots.   We then use a ``racetrack aperture" to extract satellite photometry, where the aperture radius (width perpendicular to the line connecting the spot and the star) equals that for the diffraction limit at the center wavelength for each wavelength bin (i.e. for $Y$ or $J$ band).   The aperture radial elongation is determined empirically using the start and stop wavelengths of the bin.  Photometric errors consider the intrinsic SNR of the detection, the SNR of the satellite spots, and flat-fielding errors.

   \begin{figure*}[ht!]
    \centering
    \includegraphics[width=0.99\textwidth]{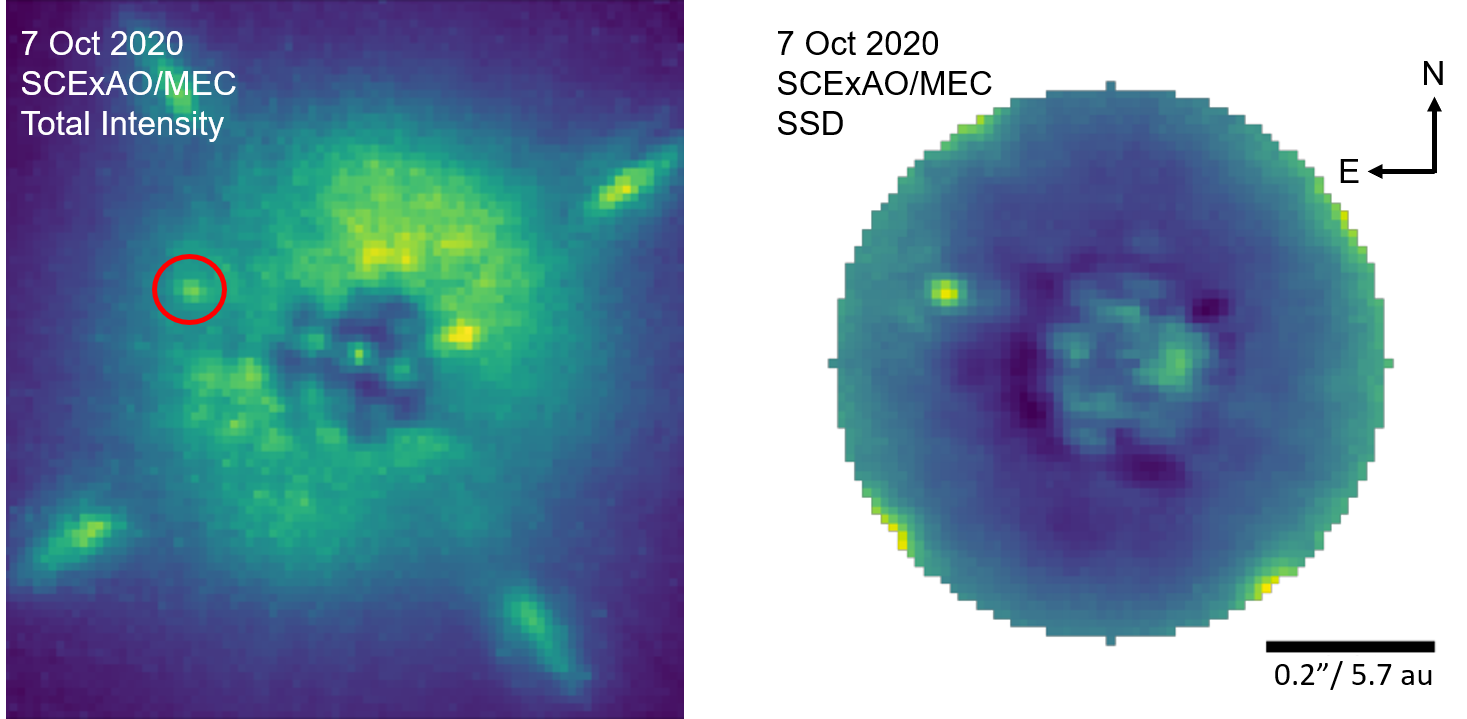}
    \caption{Left: Total intensity image of HIP 109427 B taken with SCExAO/MEC at $Y$ and $J$ band where the location of the companion has been circled in red. Right: SSD $I_C$/$I_S$ image of HIP 109427 B. Here the companion is plainly visible as well as dark regions at the edge of the coronagraph showing the removal of pinned speckles from the total intensity image.}
    \label{fig:ic_div_is}
\end{figure*}

 \subsubsection{Stochastic Speckle Discrimination (SSD) Analysis}\label{sec:SSD}

Stochastic Speckle Discrimination (SSD) is a post-processing technique first demonstrated by \citet{gladysz2008} that relies solely on photon arrival time statistics to distinguish between speckles and faint companions in coronagraphic images. 

Originally derived by \cite{goodman1975statistical}, and experimentally verified by \cite{cagigal2001} and \cite{fitzgerald2006speckle}, the underlying probability density function that estimates the intensity distribution of off-axis stellar speckles in the image plane can be given by a modified Rician (MR)

\begin{equation}
    p_{MR}(I) = \frac{1}{I_S}\exp\left({-\frac{I + I_C}{I_S}}\right) I_0 \left(\frac{2\sqrt{II_C}}{I_S}\right) 
\end{equation}

where $I_0(x)$ denotes the zero-order modified Bessel function of the first kind, $I_C$ describes the coherent intensity component attributed to the unaberrated PSF of the primary,  and $I_S$ is the time variable component of the total intensity that describes the speckle field \citep[see also][]{Marois2008b}.

For a sequence of exposures shorter than the decorrelation time of atmospheric speckles ($\sim$ 10 ms), a histogram of the image plane intensity follows a MR: $I_C$ and $I_S$ determined for each pixel in an image \citep{fitzgerald2006speckle}.   Because MEC stores the arrival time information of every photon, all time binning can be done in post-processing, which is important since the bin size that ideally samples the MR distribution is difficult to determine a priori and may vary across the image.

In order to ideally sample the MR, a bin size should be chosen that is shorter than the decorrelation timescale of the speckles in the image. If too large of a bin size is chosen, many realizations of the speckle intensity will be averaged over. Conversely, if too small of a bin size is selected, then not enough photons will arrive per bin and the distribution will tend towards Poisson statistics.

While the individual components of the MR distribution themselves do not inherently describe the signal from a faint companion, the \textit{ratio} of the coherent component to time variable component, $I_C$/$I_S$, may reveal faint companions from a comparably bright speckle field \citep{gladysz2009, meeker2018darkness}. This is because the addition of light from a companion will make the MR more negatively skewed or, analogously, increase the best fit $I_C$. This results in a larger $I_C$/$I_S$ ratio at the location of the companion compared to other pixels at the same angular separation from the primary.

We wrote an SSD analysis code to interface with the MKID Pipeline, which breaks up a MEC observation into a series of short-exposure images.  Given a user-defined bin size, we then fit a MR distribution to the histogram of the intensities for each pixel using a maximum likelihood approach. Detector dithers mitigated the large number of dead pixels in the current (engineering grade) MEC array. The SSD code is run on a single dither position at a time, and the resulting $I_C$ and $I_S$ images are drizzled together into a combined image using an adaptation of the STScI DrizzlePac software package \citep{gonzaga2012drizzlepac}.

We used this SSD code to process our 15 minute observation of HIP 109427 taken on 7 October 2020 to generate the image in Figure \ref{fig:ic_div_is}. The companion is clearly visible. Dark circular regions close to the edge of the coronagraph represent pinned speckles that have been suppressed by SSD due to their large $I_S$ component.

For this analysis, a conservative bin size of 10 ms was chosen. \citet{macintosh2005} found that speckles evolve on timescales similar to the aperture clearing time of the telescope which is given by $\tau_0 = (0.6 * D)/\overline{v}$. Here, D is the diameter of the telescope and $\overline{v}$ is the mean wind speed for the observation. During the MEC observations of HIP 109427 B, we had quite slow wind speeds of $\sim$ 5 $m/s$ which, combined with a telescope diameter of 8.2 $m$ for Subaru, yields a $\tau_0$ of $\sim$ 1 $s$. 10 $ms$ is therefore a conservative choice since we are unlikely to be sampling over more than one realization of the speckle intensity while still having enough photons per bin to not become Poissonian.

To quantify the power of this technique, we calculated the SNR by performing aperture photometry on the companion and at a series of sky apertures located in a ring at the same angular separation from the host star. These apertures all had a diameter equal to the diffraction limit at the center of the MEC bandwidth. Since the satellite spots are at a sufficient distance away from the close-in companion, all apertures were able to be used. The noise was calculated by taking the standard deviation of the sums of the sky subtracted flux for each non-companion containing aperture \citep[see also][]{Currie2011, Mawet2014}. This procedure was performed for both the total intensity and SSD $I_C$/$I_S$ image of HIP 109427 B. The SNR of the $I_C$/$I_S$ image is 21.2, about a factor of 3 higher than the SNR of 6.9 found for the total intensity image.

\begin{figure*}
    \centering
    \includegraphics[width=0.99\textwidth]{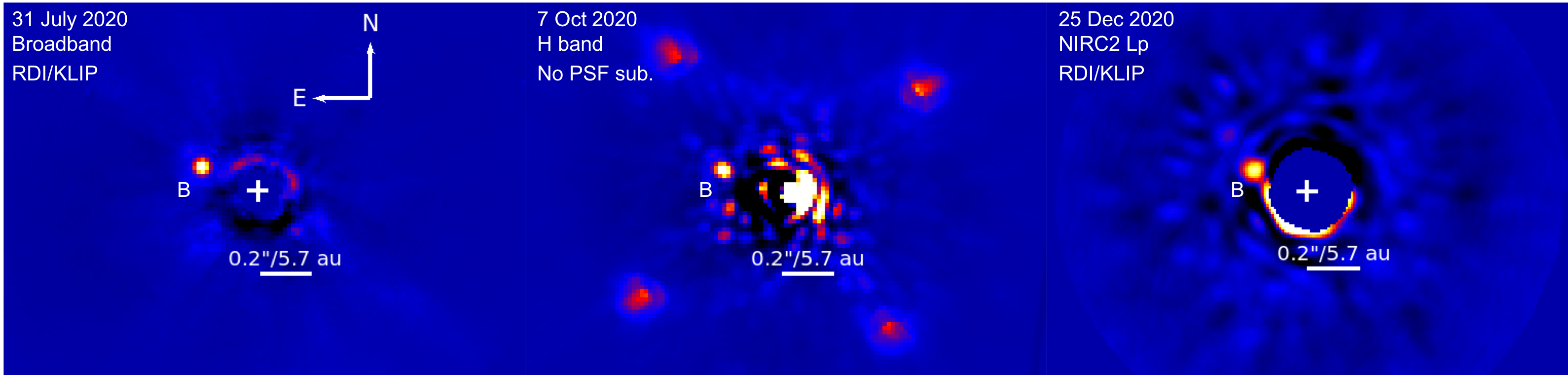}
    \vspace{-0.1in}
    \caption{Detections of HIP 109427 B from SCExAO/CHARIS in broadband ($JHK$) and $H$ band and Keck/NIRC2 in $L_{\rm p}$.   For the CHARIS broadband data (NIRC2 $L_{\rm p}$ data), we retained 5 (3) KL modes for PSF subtraction but obtain similar results for other settings.}
    \label{fig:charisnirc2}
\end{figure*}

\subsection{Image Processing: CHARIS and NIRC2}
 We extracted CHARIS data cubes from the raw data using the standard CHARIS pipeline \citep{Brandt2017} to perform basic reduction steps -- sky subtraction, image registration, and spectrophotometric calibration. For spectrophotometric calibration, we adopted a Kurucz stellar atmosphere model appropriate for an A1V star. For NIRC2 data, a well-tested general purpose high-contrast ADI broadband imaging pipeline \citep{Currie2011} performed basic processing.      To subtract the PSF for CHARIS broadband data and December 2020 NIRC2 $L_{\rm p}$ data, we used a full-frame implementation of reference star differential imaging (RDI) using the \textit{Karhunen-Loe`ve Image Projection} \citep[KLIP;][]{Soummer2012} algorithm as in \citet{Currie2019a}, although results obtained with A-LOCI were similar \citep{Currie2012,Currie2015}.   For the 2015 NIRC2 data, we used a full-frame version of A-LOCI.  

 Figure \ref{fig:charisnirc2} shows detections of HIP 109427 B in each 2020 data set.   The SNRs of HIP 109427 B in the CHARIS wavelength-collapsed broadband and $H$ band images and 2020 NIRC2 image are $\sim$ 19, 15, and 12, respectively.   HIP 109427 B is easily visible in each CHARIS channel.   We failed to obtain a decisive detection of HIP 109427 B in the 2015 NIRC2 data.  No other companions are seen in the field-of-view for any data set.

 \section{Analysis}
\begin{figure*}
    \includegraphics[width=0.5\textwidth,trim=7mm 0mm 42mm 160mm,clip]{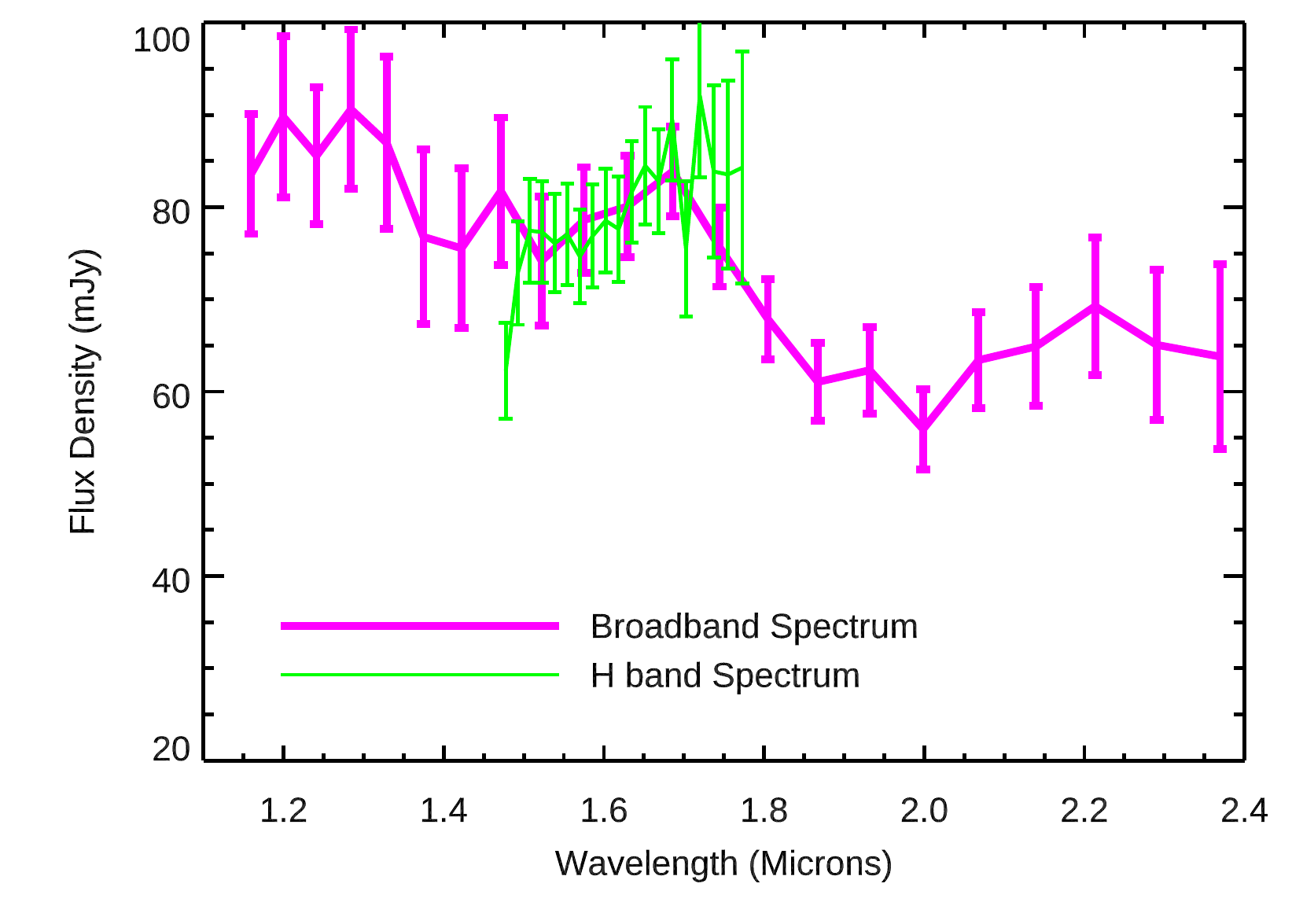}
    \includegraphics[width=0.5\textwidth,trim=0mm 5mm 0mm 0mm,clip]{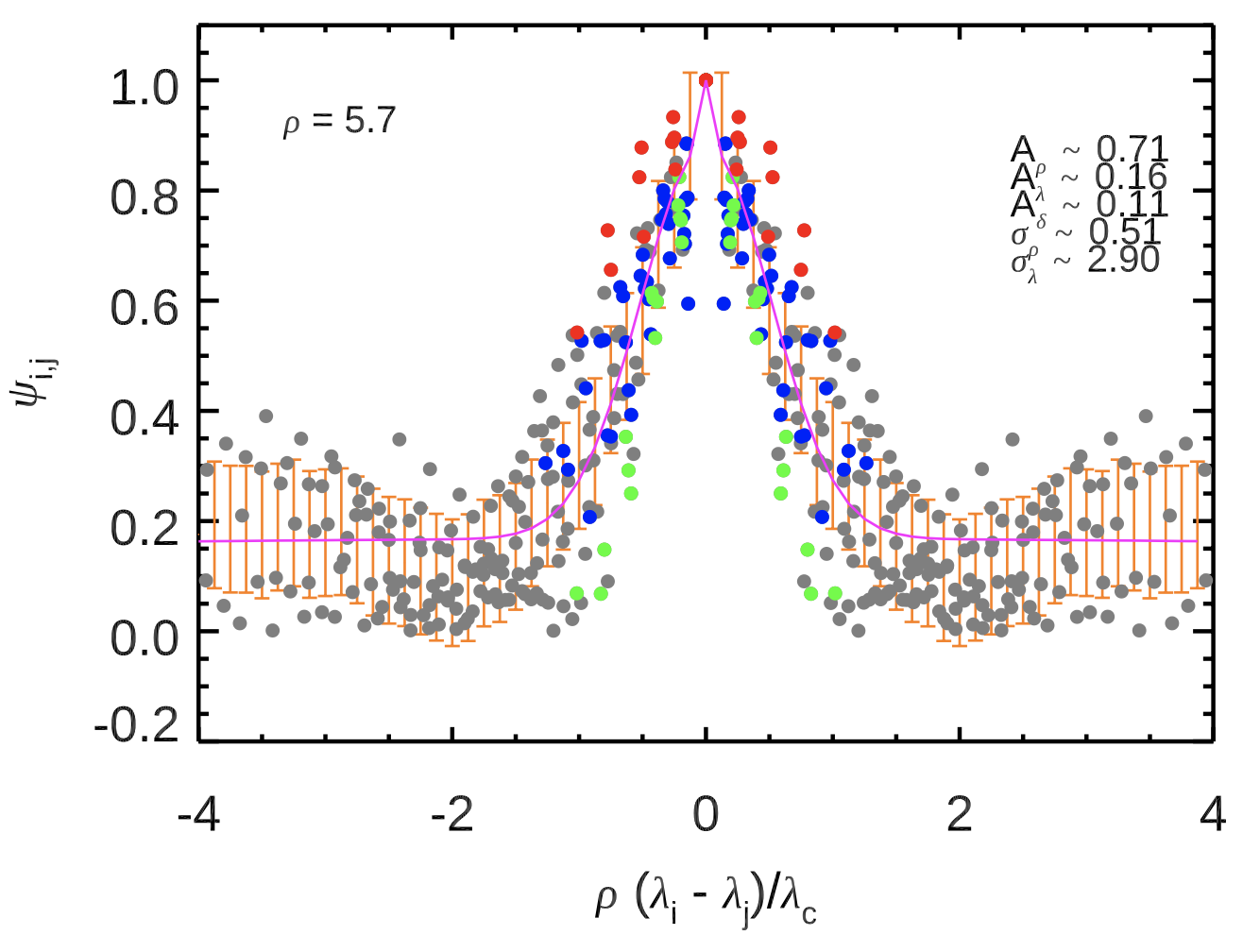}
    \caption{(left) SCExAO/CHARIS spectra for HIP 109427 B extracted from broadband data (magenta) and in $H$ band (green),(right)  Spectral covariance for the CHARIS broadband data.  The magenta line shows our fit to the spectral covariance as a function of scaled separation -- $\rho$($\lambda_{\rm i}$-$\lambda_{\rm j}$)/$\lambda_{\rm c}$ -- where $\rho$ is the separation in $\lambda$/D units for the central wavelength $\lambda_{\rm c}$ \citep[see ][]{GrecoBrandt2016}.   Blue, red, and green circles denote individual measurements between channels within the same major near-IR filter ($J$, $H$, or $K_{\rm s}$) while grey circles denote other individual measurements.   Orange points with error bars denote binned averages with 68\% confidence intervals. The broadband and $H$ band SCExAO/CHARIS data are available in the machine-readable format as data behind the Figure.}
    \label{fig:spectrum}
\end{figure*}

\subsection{HIP 109427 B Spectroscopy and Photometry}
For the CHARIS broadband data, we corrected for algorithm signal loss induced by KLIP using forward-modeling as described in \citet{Pueyo2016}.  Because we subtracted the PSF using a reference star, only oversubtraction (not self-subtraction terms) attenuates the companion signal flux and throughput is high ($\sim$95--97\%).    No throughput correction is applied for the $H$ band data since we simply subtracted a median radial profile in each channel.   The longest wavelength channel for the $H$ band spectrum was deemed unreliable due to extremely poor throughput and a large dispersion (a factor of 3) in the satellite spot flux densities used to map between counts and physical units ($mJy$).

Figure \ref{fig:spectrum} (top panel) shows the CHARIS spectrum.   The broadband and $H$ band flux densities agree to within 1-$\sigma$ except at $\sim$ 1.45~$\upmu$m, where telluric absorption is strongest. The CHARIS spectra show clear local minima at 1.4~$\upmu$m and 1.8--2.0~$\upmu$m, consistent with absorption from water opacity \citep[e.g.][]{Currie2020}.  In the standard Mauna Kea Observatory bandpasses, HIP 109427 B photometry drawn from the CHARIS broadband spectrum and NIRC2 imaging data is $J = 10.62 \pm 0.10$, $H = 10.30 \pm 0.07$, $K_{\rm s} = 10.02 \pm 0.11$, and $L_{\rm p} = 9.58 \pm 0.13$.  The MEC $Y$ and $J$ band photometry is consistent with CHARIS-drived values: $Y = 10.73 \pm 0.24$ and $J = 10.67 \pm 0.23$.

\begin{deluxetable*}{llllll}[ht]
     \tablewidth{0pt}
    \tablecaption{HIP 109427~B Detection Significance, Astrometry, and Photometry}
    \tablehead{\colhead{UT Date} & \colhead{Instrument} & \colhead{Passband} & \colhead{SNR$^{a}$} & \colhead{[E,N](\arcsec{})} & {Photometry}}
    \startdata
    20200731 &SCExAO/CHARIS & $JHK$ & 19 &  [0.229, 0.100] $\pm$ [0.004, 0.004] & $J = 10.62 \pm 0.10$, $H = 10.31 \pm 0.08$, $K_{\rm s} = 10.02 \pm 0.10$\\
    20201007 &SCExAO/MEC & $YJ$ & 7.0, 21.4$^b$ & [0.228, 0.092] $\pm$ [0.010, 0.010] & $Y = 10.73 \pm 0.23 $, $J = 10.67 \pm 0.24 $\\
    -- &SCExAO/CHARIS & $H$ & 15 & [0.229, 0.086] $\pm$ [0.004, 0.004] & $H$ = 10.28 $\pm$ 0.09\\
    20201225 &Keck/NIRC2 & $L_{\rm p}$ & 12 & [0.222, 0.077] $\pm$ [0.003, 0.003] & $L_{\rm p}$ = 9.58 $\pm$ 0.13\\
    \enddata
    \tablecomments{a) All HD 109427~B SNR estimates were drawn from reductions used to calculate astrometry. b) The higher SNR SSD image can be used to determine MEC astrometry only: MEC photometry is performed using the simple sequence-combined image without post-processing (SNR = 7.0).
    }
    \label{astrolog}
    \vspace{-0.5cm}
    \end{deluxetable*}

\begin{figure*}
    \includegraphics[width=0.5\textwidth,trim=0mm 0mm 0mm 0mm,angle=0,clip]{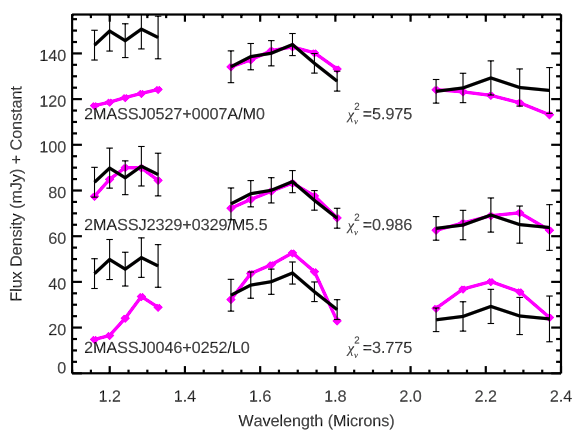}
    \includegraphics[width=0.52\textwidth,trim=0mm 0mm 0mm 0mm,clip]{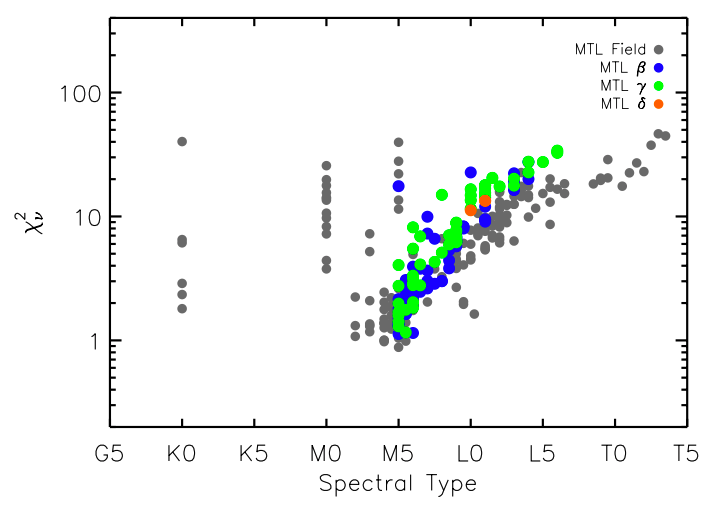}
    \vspace{-0.25in}
    \caption{(Left) The CHARIS HIP 109427 B spectrum (black) compared to field brown dwarf spectra (magenta) with M0, M5.5, and L0 spectral types from the Montreal Spectral Library (binned to CHARIS's resolution).  (Right) The $\chi^{2}_{\nu}$ distribution comparing HIP 109427 B's spectrum to objects in the Montreal Spectral Library. The blue, green, and orange symbols refer to dwarfs with gravity classifications of intermediate, low, and very low which tend to be intermediate aged (e.g. $\sim$ 100 Myr), young (10--100 Myr), and very young ($<$ 10 Myr), whereas field (older) dwarfs are shown as grey circles \citep[see ][]{Currie2018}.} 
    \label{fig:specempfit}
\end{figure*}

\subsection{HIP 109427 B Spectral Type, Temperature, and Luminosity}
Following recent work \citep{Currie2020}, we compared the CHARIS spectra for HIP 109427 B to entries in the Montreal Spectral Library\footnote{\url{https://jgagneastro.com/the-montreal-spectral-library/}} \citep[e.g.][]{Gagne2015}, considering the impact of spatially and spectrally correlated noise \citep{GrecoBrandt2016}\footnote{We do not also compare the MEC or NIRC2 photometry due to sparse coverage of the library outside of the $JHK$ passbands}.   The CHARIS data reveal highly correlated errors (Figure \ref{fig:spectrum}, right panel).  The spectral covariance at HD 109427 B's location includes substantial off-diagonal terms, especially for spatially-correlated noise (A$_{\rm \rho}$ $\sim$ 0.71) and (to a lesser extent) residuals speckles well correlated as a function of wavelength (A$_{\rm \lambda}$ $\sim$ 0.16). 

As shown in Figure \ref{fig:specempfit}, HIP 109427 B's CHARIS spectrum is best matched by M4--M5.5 field objects (left panel).   Three objects in the Montreal library yield $\chi_{\nu}^{2}$ $\le$ 1, even with the full spectral covariance included: 2MASSJ0326-0617 (M5), 2MASSJ0854-3051 (M4), and 2MASSJ2329+032 (M5.5).   Using the mapping between spectral type and effective temperature from \citet{Pecaut2013}, empirical comparisons to the CHARIS spectra then favor a temperature of 3000-3200 $K$ for HIP 109427 B.  Adopting the relationship from \citet{Casagrande2008} and assuming a distance of 28.3 $pc$, HIP 109427 B's luminosity is $\log_{10}(L/L_{\rm \odot}) = -2.28^{+0.04}_{-0.04}$.

We compared the MEC $YJ$-band photometry, CHARIS $JHK$ spectra, and NIRC2 $L_{\rm p}$ photometry to the BT-Settl atmosphere models \citep{Allard2012} with the \citet{Asplund2009} abundances downloaded from the Theoretical Spectra Web Server\footnote{\url{http://svo2.cab.inta-csic.es/theory/newov2/}}.    The grid covers temperatures of 2500--4000 $K$, surface gravities of log(g) = 3.5--5.5, and metallicities of [Fe/H] = -1 to 0.5.   Following \citet{Currie2018b}, we focus only on the CHARIS channels unaffected by telluric absorption, resulting in 21 photometric/spectrophotometric points fit.   We define the fit quality for the $kth$ model using the $\chi^{2}$ statistic, considering the spectral covariance:
\begin{equation}
    \chi^{2} = R_{k}^{T}C^{-1}R_{k} + \sum_{i}(f_{phot,i}-\alpha_{k}~F_{phot,ik})^{2}/\sigma_{phot,i}^{2}.
\end{equation}
Here, the vector $R_{k}$ is the difference between measured and predicted CHARIS data points ($f_{spec}-\alpha_{k}F_{spec}$) and $C$ is the covariance for the CHARIS spectra.  The vectors $f_{phot,i}$, $F_{phot,ik}$, and $\sigma_{phot,i}$ are measured photometry, model predicted photometry, and photometric uncertainty; $\alpha_{k}$ is the scaling factor for the model that minimizes $\chi^{2}$ \citep[see also][]{DeRosa2016}.

\begin{figure*}
    \centering
    \includegraphics[width=0.95\textwidth,trim=10mm 4mm 0mm 5mm,angle=0,clip]{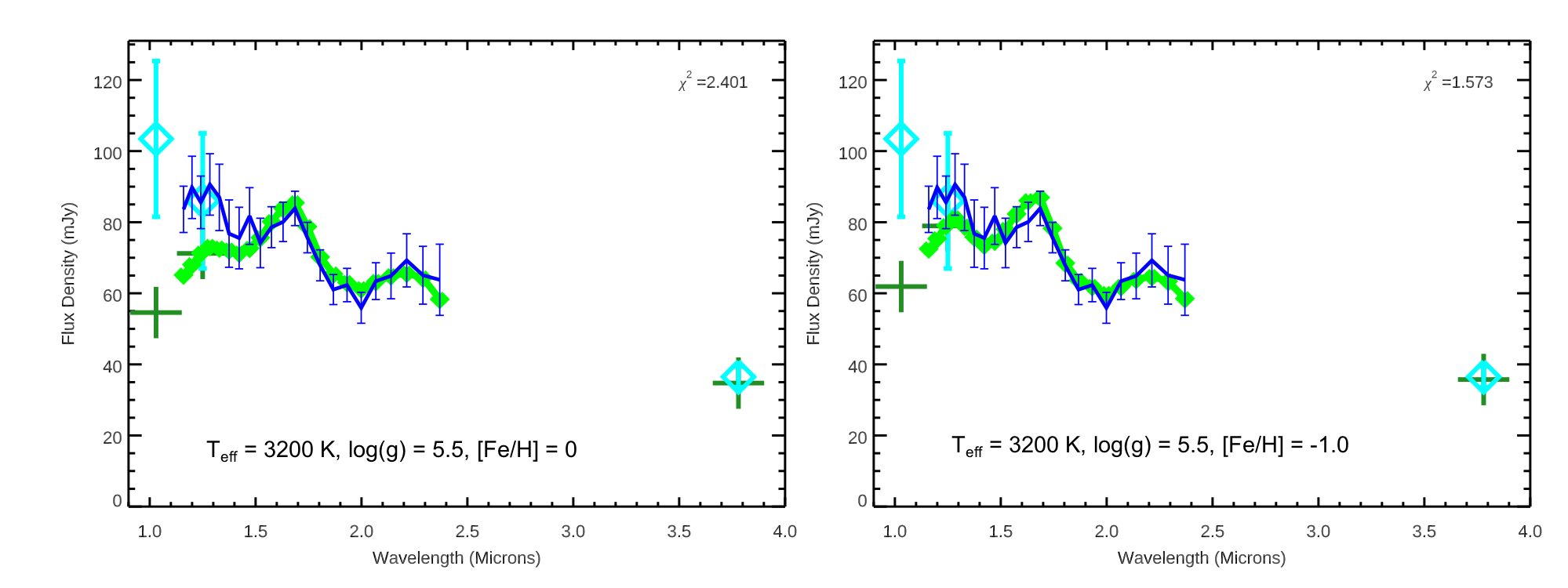}
    \includegraphics[width=0.45\textwidth,trim=-6mm 0mm 20mm 5mm,clip]{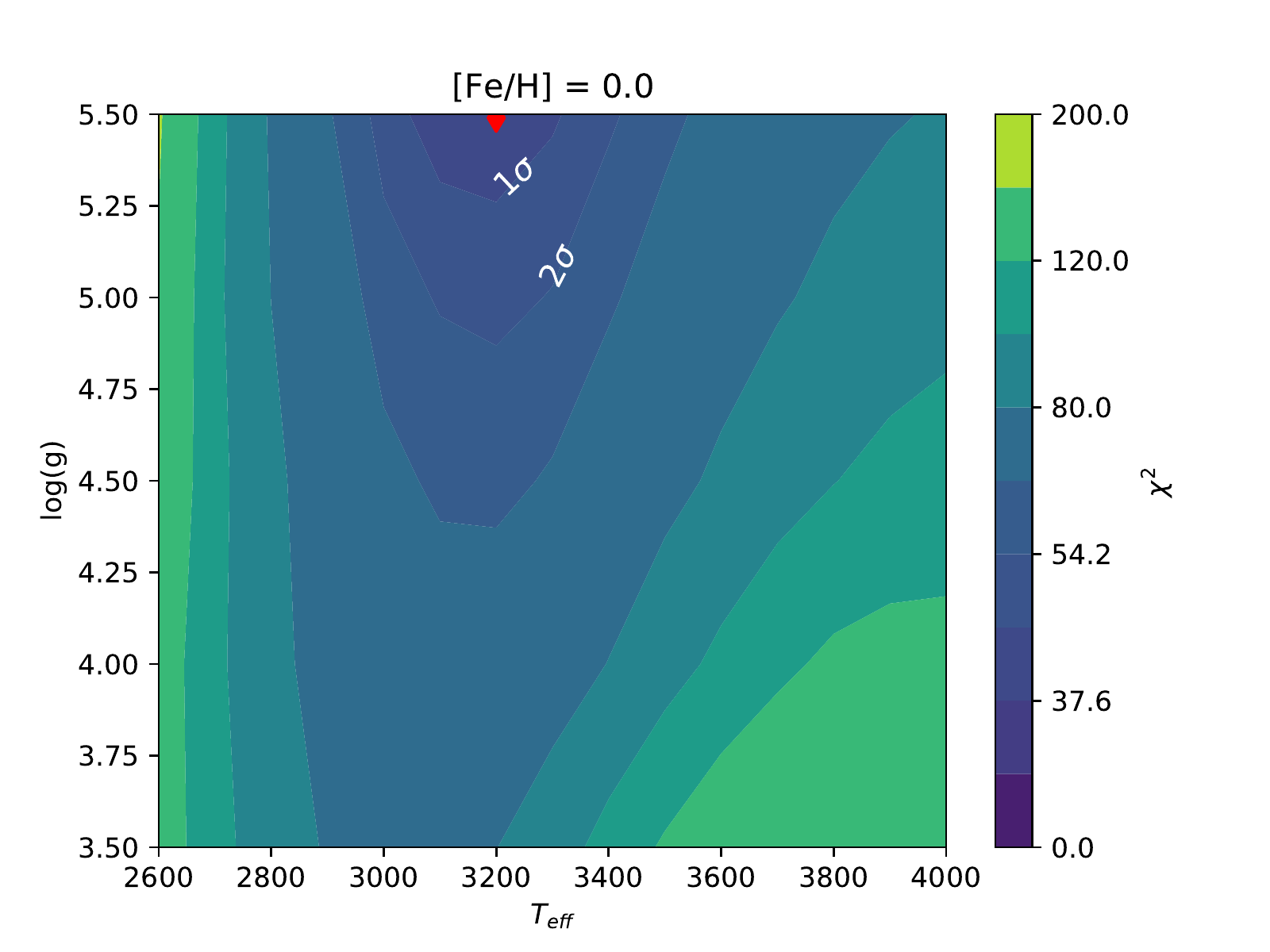}
    \includegraphics[width=0.45\textwidth,trim=-6mm 0mm 20mm 5mm,clip]{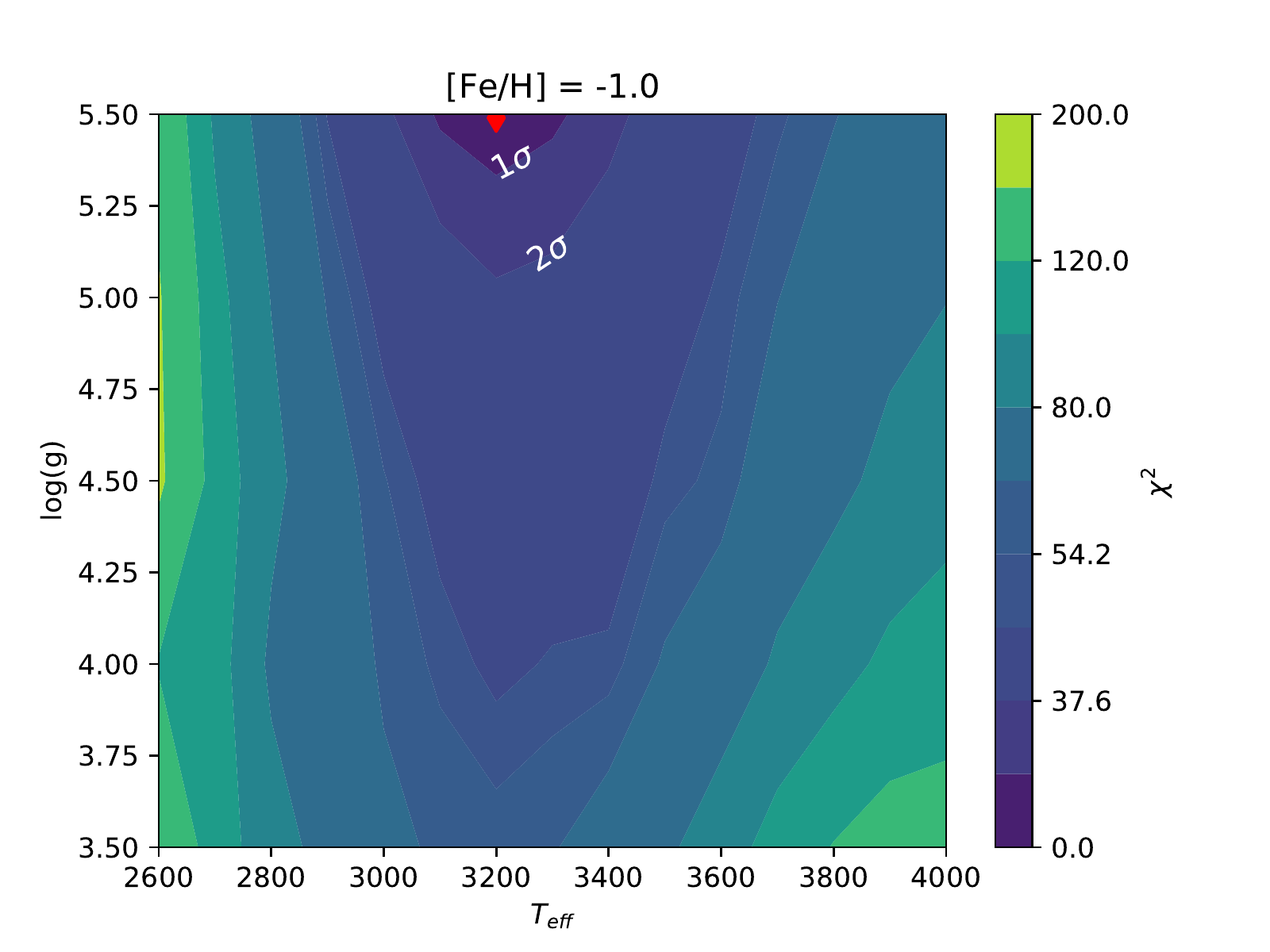}
    \vspace{-0.05in}
    \caption{(top) Best fit BT-Settl models for a solar and non-solar metallicity and (bottom) corresponding contour plots of $\chi^{2}$ as a function of temperature and surface gravity. The 1$\sigma$ and 2$\sigma$ contours are labeled in white and the best fit solution denoted with a red diamond.  The $\chi_{\nu}^{2}$ value shown is for 20 degrees of freedom.   CHARIS spectra is shown in blue, MEC and NIRC2 photometry in cyan, model-predicted CHARIS spectrophotometry in light green, and predicted MEC/NIRC2 photometry in dark green.}
    
    \label{fig:btsettlfit}
\end{figure*}

Figure \ref{fig:btsettlfit} shows the best-fit solar and non-solar metallicity models (top panels) and the associated $\chi^{2}$ contours (bottom panels).   An atmosphere with a temperature of $T_{\rm eff}$ = 3200 $K$ and a high gravity (log(g) = 5.5) fits the data the best in both cases.   The 1-$\sigma$ contour for temperature and gravity is narrowly defined about this peak for both metallicities: $T_{\rm eff}$ = 3100--3300 $K$ and log(g) = 5.25--5.5.   At the 2-$\sigma$ level, the best-fit temperature and gravity ranges widen to 3000--3400 $K$ and log(g) = 5--5.5.   The radii that minimize $\chi^{2}$ are $\sim$2.1--2.6 Jupiter radii.

The best-fit solar metallicity model accurately reproduces the $H$ and $K$ portions of CHARIS spectrum and the NIRC2 $L_{\rm p}$ photometry; however, it underpredicts the brightness of HIP 109427 B in $Y$ and $J$ band by 85\% and 25\%, respectively.   Subsolar metallicity models systematically produce a rough match in $J$ band and show less severe disagreement at $Y$ band.   Future MEC calibration work, such as improving the wavelength dependent flat-fielding, may yield better agreement with expected $Y$ band photometry.

The 2-$\sigma$ ranges for temperature correspond to M3--M5.5 dwarfs, a range that overlaps with the spectral types of best-matching objects in the Montreal Spectral Library, although the best-fit is skewed towards earlier, hotter objects by by $\sim$1 subclass.  For M3--M5.5 objects with the HIP 109427 system's estimated age of $\sim$ 0.4--0.7 $Gyr$, the expected surface gravities are log(g) $\sim$ 5--5.1 \citep{Baraffe2003}, or about 0.25--0.5 dex lower than the best-fit values considered by our grid.   Expected radii are 2--3 Jupiter radii: consistent with our best-fit values.

    \begin{figure}
    \centering
    \includegraphics[width=0.87\columnwidth, trim=-8mm 0mm 0mm 0mm,clip]{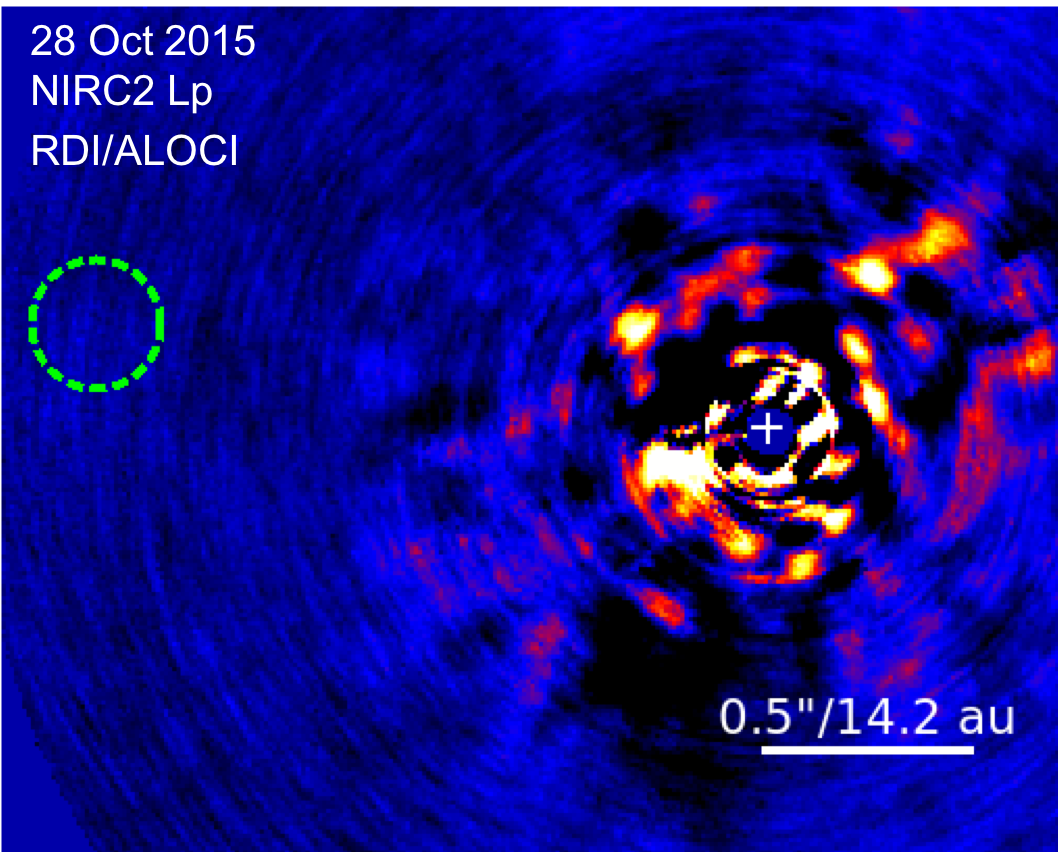}
    \includegraphics[width=\columnwidth,trim=15mm 5mm 0mm 5mm,clip]{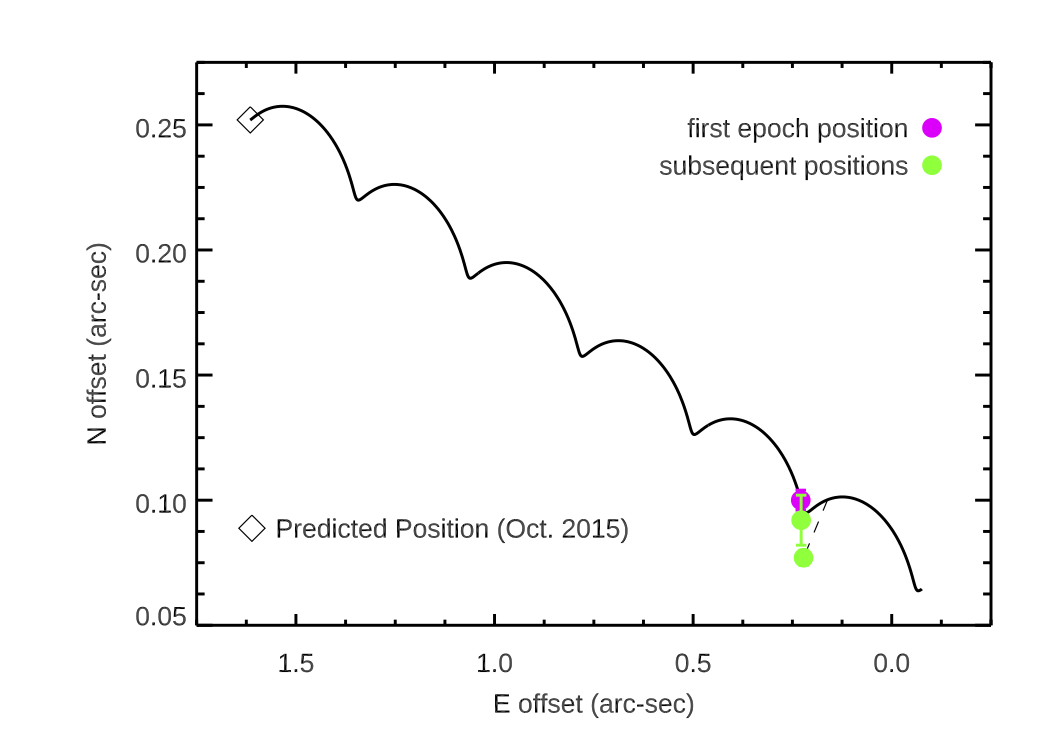}
    \caption{(top) Keck/NIRC2 data taken in $L_p$ showing a non-detection at the expected location of the companion, which is circled in green. (bottom) Expected track for a background object showing its predicted location in October 2015.  The dashed line connects the measured Dec 2020 position to the predicted position for a background object.}
    \label{fig:nirc2archive}
\end{figure}

\subsection{HIP 109427 B Astrometry and Dynamical Mass}

\subsubsection{Evidence for Common Proper Motion}

To rule out the possibility that HIP 109427 B is a background object, we analyzed archival 2015 Keck/NIRC2 data shown in Figure \ref{fig:nirc2archive}.    The data do not reveal a statistically significant detection of any signal that could be HIP 109427 B. Using the small sample statistics correction from \citet{Mawet2014}, we estimate a 5-$\sigma$ contrast of $\Delta L_{\rm p}$ $\sim$ 5, 5.75, 11.3, and 12 magnitudes at 0\farcs{}15, 0\farcs{}225, 1\farcs{}0, and 1\farcs{}5, respectively. Companions at HIP 109427 B's current angular separation would be just undetectable at 5-$\sigma$.  Those with contrasts like HIP 109427 B near 2 $\lambda$/D would be well below the detection limit and those at arcsecond or wider separations would be easily detected.

HIP 109427 has an extremely high proper motion of $\mu_{\rm \alpha}~cos(\delta)$, $\mu_{\rm \delta}$ $\sim$ 282.18, 30.46 mas yr$^{-1}$  \citep{vanLeeuwen2007}. If HIP 109427 B were a background star, it would appear at an angular separation of $\sim$1\farcs{}6 in October 2015 data with an expected SNR of $\sim$ 1000.   However, no signal is present at its expected location (dashed circle).    HIP 109427 B's position in December 2020 also deviates by $\sim$ 65 mas compared to the expected location of a background star, far larger than our astrometric precision (bottom panel).   This implies HIP 109427 B is a common proper motion companion to the primary.


\subsubsection{Orbit and Dynamical Mass}
We used the open source code \verb|orvara|, \citet{brandt_orvara}, to fit the mass and orbit of HIP 109427 B.  \verb|orvara| is an MCMC orbit fitting code for exoplanetary systems that uses a combination of absolute astrometry, relative astrometry, and radial velocities (RVs) to fit one or more Keplerian orbits to a system. For this dataset, we used HGCA absolute astrometry measurements for the star and the three measured epochs of relative astrometry for the companion from CHARIS and MEC. We do not consider RV limits since previous data has had a limited time baseline and poor precision.   A Gaussian prior of $2.1 \pm 0.15$ $M_\odot$ was chosen for the primary in concordance with literature values derived from isochrone fitting \citep{vast2014, David2015, Stone2018}.

\begin{figure*}
    \centering
    \includegraphics[width=\textwidth]{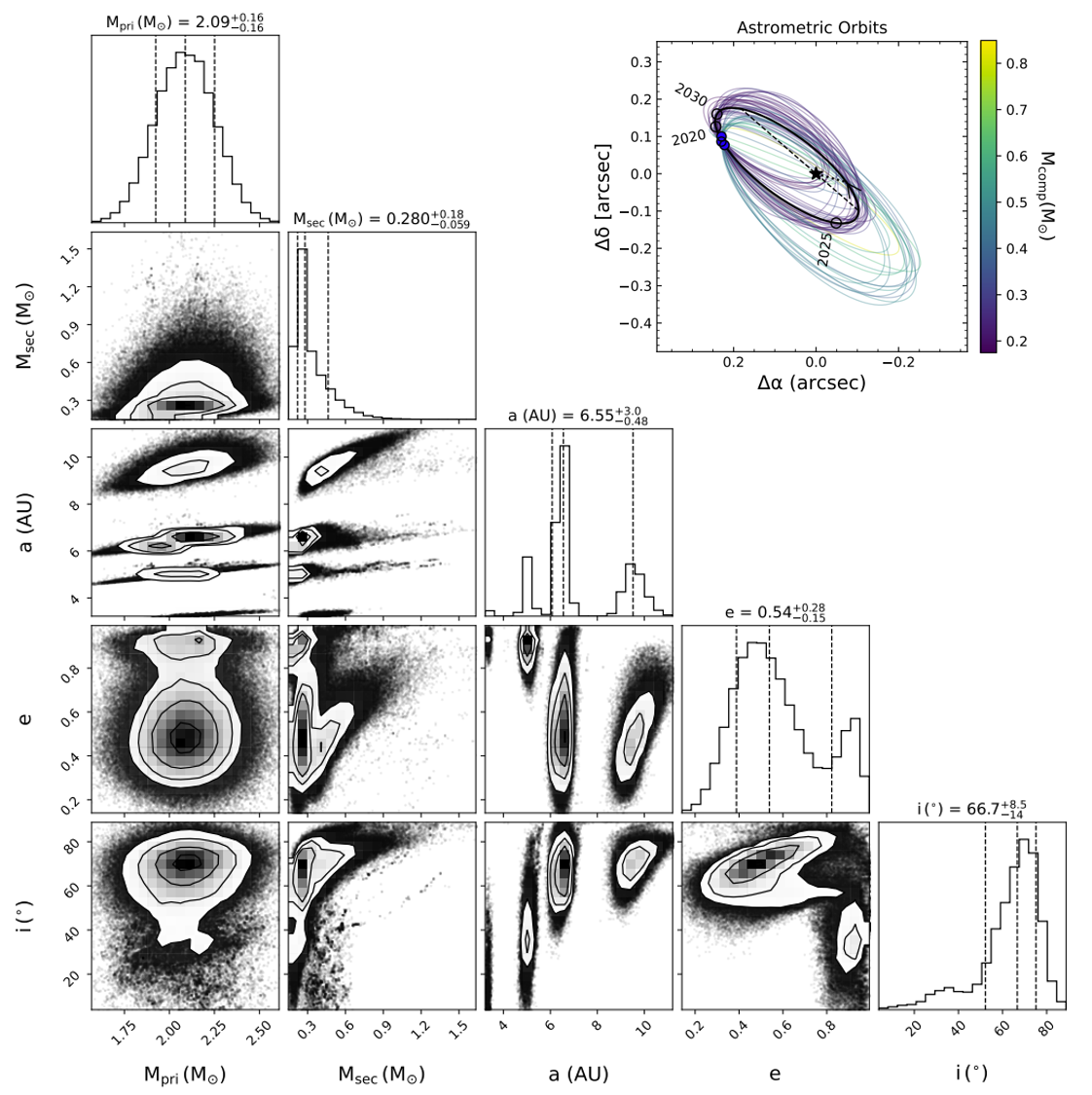}
    \caption{Corner plot displaying select posterior orbital parameters. The orbit fits were performed using HGCA data and relative astrometry points from SCExAO/CHARIS and MEC data. The mass of the primary is nearly identical to the chosen prior of $2.1^{+0.15}_{-0.15}$ $M_{\odot}$. (Inset) The best fit orbit of HIP 109427 B in black with 50 randomly selected orbits from the MCMC fit color-coded by HIP 109427 B's mass. The blue circles represent the measured relative astrometry points and the unfilled black circles are the predicted locations of the companion at different epochs. The arrow indicates that HIP 109427 B is orbiting counter-clockwise}
    \label{fig:orbit_fit}
\end{figure*}

Figure \ref{fig:orbit_fit} shows the posterior distributions of select orbital parameters as well as the primary and secondary mass. A summary of the fit parameters can also be found in Table \ref{orbit_params}. The mass of the primary is nearly identical to the adopted prior with a value of $2.09 ^{+0.16}_{-0.16}$ $M_\odot$ and the fit secondary mass is $0.280^{+0.18}_{-0.059}$ $M_\odot$. The best fit eccentricity is $0.54^{+0.28}_{-0.15}$ with an inclination of $66.7^{+8.5}_{-14}$ degrees. The best fit semimajor axis is $6.55^{+3.0}_{-0.48}$ au, although the distribution is bimodal with HIP 109427 B's mass with one family of solutions favoring a $\sim$6 au separation with a mass of $\sim$ 0.25 $\pm$ 0.05 $M_{\odot}$ and another favoring a mass of 0.5 $M_{\odot}$ and semimajor axis of 9 au.   Main-sequence stars with masses of 0.5 $M_{\odot}$ have early M spectral types \citep[e.g.][]{Pecaut2013}, which are excluded from our spectral analysis.   In contrast, the lower-mass solution is consistent with M4 V object allowed by the CHARIS spectral comparisons.   

A mass of $\sim$0.25 $M_{\odot}$ is broadly consistent with inferred masses based on luminosity evolution models, given HIP 109427 B's likely age.  From the \citet{Baraffe2003} models, an M3--M5.5 object with an age of 400--700 $Myr$ is predicted to have a mass of 0.15--0.3 $M_{\odot}$.   Modeling absolute astrometry of the primary and relative astrometry of the star likely then yields much more precise (20\%) constraints on the companion mass than available from luminosity evolution models alone (50\%).

\begin{deluxetable}{llllllllll}
    \tablewidth{\columnwidth}
    \tabletypesize{\small}
    \tablecaption{HIP 109427 B Orbit Fitting Results and Priors}
    \tablehead{\colhead{Parameter} & \colhead{Fitted Value} &  \colhead{Prior}}
    \startdata
    $M_{pri}$ ($M_{\odot})$ & $2.09 \pm 0.16$ & Gaussian,  $2.1 \pm 0.15$\\
    $M_{sec}$ ($M_{\odot})$ & $0.280^{+0.18}_{-0.059}$ & 1/$M_{sec}$ \\
    Semimajor axis \textit{a} (au) & $6.55^{+3.0}_{-0.48}$ & 1/a \\
    Eccentricity \textit{e} & $0.54^{+0.28}_{-0.15}$ & uniform \\
    Inclination \textit{i} ($^{\circ}$) & $66.7^{+8.5}_{-14}$ & sin(\textit{i})\\
    \enddata
    \tablecomments{Posterior distributions for the secondary mass and semimajor axis are bimodal with a favored solution of $\sim$ 0.25 $M_{\odot}$ and $\sim$ 6 au - see Figure \ref{fig:orbit_fit} and text for more details.}
    \label{orbit_params}
    \vspace{-1.0cm}
\end{deluxetable}

\section{Conclusion}

With SCExAO/MEC photometry, SCExAO/CHARIS spectroscopy, and Keck/NIRC2 photometry, we have identified a low mass stellar companion at a near-Jupiter-like separation around the nearby A1V star HIP 109427. Comparison of this target's spectrum with entries in the Montreal Spectral Library indicates a spectral type of M4--M5.5. This is consistent with a best fit a dynamical mass of $\sim 0.25$ $M_{\odot}$ with a semimajor axis of $\sim 6$ au from orbital fitting using measurements from both \textit{Hipparcos} and \textit{Gaia} DR2 as well as MEC, CHARIS, and NIRC2 relative astrometry. There is a degeneracy in the orbital fit with another favored solution of $\sim 0.5$ $M_{\odot}$ with a semimajor axis of $\sim 9$ au that is excluded by our spectral analysis. Future RV measurements, Gaia astrometry, and relative astrometry from high-contrast imaging will help to better constrain this orbit.

This result demonstrates the efficacy of Stochastic Speckle Discrimination (SSD) in identifying faint companions.  SSD increases the SNR of HIP 109427 B by about a factor of 3 versus the total intensity image (comparable to the CHARIS SNR of this target) without the use of any additional PSF subtraction techniques. This technique is especially effective at small angular separations (inside 10 $\lambda$/D) where algorithms exploiting traditional observing strategies like ADI and SDI suffer.

Work expanding the SSD framework to be agnostic to bin size and to directly fit an off-axis Poisson source has been shown to be effective on simulated data and is currently being adapted for use on real datasets \citep{walter2019stochastic}. Current hurdles in adapting this technique likely stem from key differences between the simulated dataset and real on-sky data. Specifically, we are exploring the effects of variable Strehl during an observation and speckle chromaticity. Once effectively adapted to real data, this new bin-free SSD technique will allow us to not only remove bin size as a variable in our analysis, but also to directly extract the component of the intensity attributable to the companion itself. Unlike the $Ic/Is$ maps in this work, which are limited to highlighting regions of an image that contain a companion of comparable brightness to the surrounding speckle field, this information could then be fed into other traditional post-processing techniques (such as ADI and SDI) to further improve the SNR of faint companion detections and help image companions buried beneath the speckle noise.

\newpage
\newpage

\section{Acknowledgements}
S.S. and N.F. are supported by a grant from the Heising-Simons Foundation. K.K.D. is supported by the NSF Astronomy and Astrophysics Postdoctoral Fellowship program under award \# 1801983.  J.P.S. and N.Z. are both supported by a NASA Space Technology Research Fellowship. I.L. is supported by the National Science Foundation Graduate Research Fellowship under grant \# 1650114.  T.C. was supported by a NASA Senior Postdoctoral Fellowship and NASA/Keck grant LK-2663-948181. M.T. is supported by JSPS KAKENHI grant \# 18H05442, \# 15H02063, and \# 22000005. \\

We thank the Subaru and NASA Keck Time Allocation Committees for their generous support of this program and Chas Beichman and Dawn Gelino for graciously supporting additional NIRC2 time.    This work makes use of the Keck Observatory Archive (KOA), a joint development between the W. M. Keck Observatory and the NASA Exoplanet Science Institute (NExScI).     
CHARIS was built at Princeton University under a Grant-in-Aid for Scientific Research on Innovative Areas from MEXT of the Japanese government (\# 23103002)\\

The development of SCExAO was supported by the Japan Society for the Promotion of Science (Grant-in-Aid for Research \#23340051, \#26220704, \#23103002, \#19H00703 \& \#19H00695), the Astrobiology Center of the National Institutes of Natural Sciences, Japan, the Mt Cuba Foundation and the director’s contingency fund at Subaru Telescope. The authors wish to recognize and acknowledge the very significant cultural role and reverence that the summit of Maunakea has always had within the indigenous Hawaiian community, and are most fortunate to have the opportunity to conduct observations from this mountain.

\clearpage

\bibliography{hip109427b}{}
\bibliographystyle{aasjournal}



\end{document}